\documentclass[useAMS,usenatbib]{aa}

\usepackage{graphicx}
\usepackage{amsmath}
\usepackage{txfonts}
\usepackage{fixltx2e}

\usepackage{multirow}
\usepackage[sort,round]{natbib}
\usepackage{placeins}
\usepackage{url}

\newcommand{\xmm}{{XMM-{\em Newton} }}
\newcommand{\xmmns}{{XMM-{\em Newton}}}
\newcommand{\swift}{{\em Swift }}
\newcommand{\ngswift}{Neil Gehrels Swift Observatory }
\newcommand{\swiftns}{{\em Swift}}
\newcommand{\rosat}{{\em ROSAT }}
\newcommand{\srcnamelong}{{XMMSL2~J144605.0+685735 }}

\newcommand{\srcname}{{XMMSL2~J1446+68 }}
\newcommand{\srcnamens}{{XMMSL2~J1446+68}}
\newcommand{\srcnamemlong}{{2MASX~14460522+6857311 }}
\newcommand{\srcnamemlongns}{{2MASX~14460522+6857311}}
\newcommand{\srcnamem}{{2MASX~1446+6857 }}
\newcommand{\srcnamemns}{{2MASX~1446+6857}}
\newcommand{\mseven}{{2MASX~0740-85 }}
\newcommand{\msevenns}{{2MASX~0740-85}}

\newcommand{\sdss}{{SDSS~J1201+30 }}
\newcommand{\swtd}{{SWIFT~J164449.3+573451 }}

\newcommand{\assasn}{{ASSASN-14li }}

\newcommand{\fluxunits}{{ergs s$^{-1}$cm$^{-2}$ }}
\newcommand{\fluxUnits}{{ergs s$^{-1}$cm$^{-2}$ }}
\newcommand{\fluxUnitsns}{{ergs s$^{-1}$cm$^{-2}$}}
\newcommand{\lumunits}{{ergs s$^{-1}$ }}
\newcommand{\lumUnits}{{ergs s$^{-1}$ }}
\newcommand{\lumunitsns}{{ergs s$^{-1}$}}
\newcommand{\lumUnitsns}{{ergs s$^{-1}$}}

\newcommand{\chired}{{$\chi^{2}_{r}$}}
\newcommand{\msolar}{$M_{\odot}$ }
\newcommand{\msolarns}{$M_{\odot}$}

\begin{document}

   \title{\srcnamelong : a slow tidal disruption event 
}

   \subtitle{}

   \author{R.D. Saxton
          \inst{1}
          \and
          A.M. Read\inst{2}
          \and
          S. Komossa\inst{3}
          \and
          P. Lira\inst{4}
          \and
          K.D. Alexander\inst{5}\thanks{NHFP Einstein Fellow }
          \and
          I. Steele\inst{6}
          \and
          F. Oca{\~{n}}a\inst{7}
          \and
          E. Berger\inst{8}
          \and
          P. Blanchard\inst{8}
          }

   \offprints{R. Saxton}

   \institute{Telespazio-Vega UK for ESA, European Space Astronomy Centre, Operations Department, 28691 Villanueva de la Ca\~{n}ada, Spain\\
              \email{richard.saxton@sciops.esa.int}
         \and
             Dept. of Physics and Astronomy, University of Leicester, Leicester LE1 7RH, U.K.
         \and
             Max Planck Institut f\"ur Radioastronomie, Auf dem Huegel 69, 53121 Bonn, Germany
         \and
             Universidad de Chile, Observatorio Astronomico Nacional Cerro Calan, Santiago, Chile
         \and
             Center for Interdisciplinary Exploration and Research in Astrophysics (CIERA) and Department of Physics and Astronomy, Northwestern University, Evanston, IL 60208, USA
         \and
             Liverpool John Moores University, Liverpool, L3 5RF, UK
         \and
             Dpto. de F\'isica de la Tierra y Astrof\'isica, Facultad de Ciencias F\'isicas, Universidad Complutense de Madrid, E-28040 Madrid, Spain
         \and
             Harvard-Smithsonian Center for Astrophysics, 60 Garden St., Cambridge, MA 02138, USA
        }

   \date{Received September 15, 1996; accepted March 16, 1997}

  \abstract
   {}
   {We   investigate the evolution of X-ray selected tidal disruption events.}
   {New events are found in near real-time data from XMM-Newton slews, and are monitored
by multi-wavelength facilities.}
   {In August 2016, X-ray emission was detected
from the galaxy \srcnamelong (also known as \srcnamemlongns),  that was 20 times higher than an upper limit
from 25 years earlier. The X-ray flux was flat for $\sim100$ days and then fell by a factor of 100 over the following 500 days. The UV flux was stable for the first 400 days before
fading by a magnitude, while the optical (U,B,V) bands were roughly constant for $850$ days.
Optically, the galaxy appears to be quiescent, at a distance of $127\pm{4}$ Mpc (z=$0.029\pm{0.001}$)
with a spectrum consisting of a young stellar population of   
1-5 Gyr in age, an older population, and a total stellar mass of $\sim6\times10^{9}$\msolarns.
 The bolometric luminosity peaked at L$_{bol}\sim10^{43}$ \lumUnits
with an X-ray spectrum that may be modelled by 
a power law of $\Gamma\sim2.6$ or Comptonisation of a low-temperature thermal component by thermal electrons. 
We consider a tidal disruption event to be the most likely cause of the flare.
Radio emission was absent in this event down to $<10\mu$Jy, which limits the total 
energy of a hypothetical off-axis jet to $E<5\times10^{50}$ ergs.
The independent behaviour of the optical, UV, and X-ray light curves challenges models
where the UV emission is produced by reprocessing of thermal nuclear emission
or by stream-stream collisions. We suggest that the observed UV emission may have been produced from a truncated accretion disc and the X-rays from Compton upscattering of these disc photons.
}
   {}


   \maketitle
%

\section{Introduction}

The close approach of a star to a supermassive black hole (SMBH) can lead to the destruction 
of the stellar body in a process known as a tidal disruption event \citep[TDE;][]{Hills}.
Gravitationally bound material returns to the black hole and is accreted, 
giving rise to a flare whose electromagnetic signature peaks in the 
extreme ultraviolet (EUV) band \citep{Rees88,Ulmer99}. These flares were first detected in the soft X-ray band by ROSAT \citep{Komossa1242,Bade96,Komossa99b,KomossaBade}, later by \xmm and Chandra
\citep{Esquej07,Saxton12,Saxton17,Maksym10,Lin15} \citep[see review by ][]{Komossa17}, and also in the UV band by GALEX \citep{Gezari06,Gezari08,Gezari09}.
In recent years, large-area optical surveys have detected candidate TDEs emitting at temperatures of a 
few $\times10^{4}$ K \citep{vanVelzen11,Cenko12b,Gezari12,Arcavi14,Holoien16a}, ostensibly too cool
to be coming from an accretion disc  \citep[e.g. ][]{Bonning07}.
This optical radiation has been interpreted as being due to reprocessing of the accretion radiation
by an optically thick screen \citep{MetzStone16,Roth18,Dai18} or to emission from shocks
\citep{Piran15}. 
Super-Eddington accretion in the initial phase of the disruption  causes a large-scale, radiation-driven outflow of material from the central engine \citet{Strubbe09}, which \citet{MetzStone16} showed would
initially completely absorb the radiation from the central engine
and convert it into optical/UV photons with an effective temperature similar to that 
observed. In this model, the screen density is expected to drop after a few months to the point where the inner thermal radiation would become visible, with the delay time and
ratio of X-ray to optical/UV flux depending on the line of sight \citep{MetzStone16,Dai18}. 
Observationally, the evidence for differences in the X-ray and UV/optical timescales is mixed. The X-rays may have lagged the UV by $\sim32$ days in
ASASSN-14li \citep{Pasham17}, but broadly fell on the same timescale \citep{Brown17}, as they did in \mseven \citep{Saxton17}. In \sdss the UV flux did not change, while the X-rays dropped by a factor of 100 \citep{Saxton12}, whereas in ASASSN-15oi the X-ray luminosity was quite low
($L_X\sim~10^{41}$\lumunitsns) at the peak of the optical flare, but 200--300 days later
had increased to $L_X\sim~10^{42}$\lumunits \citep{Holoien18,Gezari17}.

Clear evidence of reprocessing is offered by broad low-ionisation optical lines, which indicate a large covering angle of material orbiting the nucleus. \citet{Komossa08} identified broad hydrogen and helium lines and highly
ionised narrow iron lines in SDSS~J095209.56+214313.3, which were seen to fade over time. Similar bright,
broad lines have been seen in other archival SDSS data \citep{Wang11} and in more recent optically discovered
TDE such as \assasn \citep{Holoien16a}. Reprocessing features are, however, not generally seen in the
post-disruption optical spectra of X-ray selected TDEs on timescales of days to months \citep[e.g.][]{Saxton12, Saxton17} or years \citep[e.g.][]{Komossa99b,Lin17}.
Hence, the relationship between the emission mechanisms in the X-ray and UV bands remains an outstanding problem. 
 
In August 2016, the \xmm slew survey \citep{Saxton08} detected a flare from the nucleus of
the quiescent galaxy \srcnamemlongns. Further monitoring has shown enhanced
emission in the UV band. In Section 2 we discuss the discovery of this flare and the source identification.
In Sections 3, 4, and  5 we present X-ray, UV, optical, and radio follow-up 
observations. In Section 6 we perform a temporal and spectral analysis
of the source, and in Section 7 we discuss the flare
characteristics within the TDE model. The paper is summarised in Section 8.

A $\Lambda$CDM cosmology with ($\Omega_{M},\Omega_{\Lambda}$) = (0.27,0.73)
and  $H_{0}$=70 km$^{-1}$s$^{-1}$ Mpc$^{-1}$ has been assumed throughout.

\section{X-ray flare identification}
During the slew 9305900002, performed on August 22, 2016, \xmm 
\citep{jansen} detected
a source, \srcnamelong (hereafter \srcnamens), with an EPIC-pn, medium filter,
 0.2--2 keV count rate of 
$1.2\pm0.4$ count s$^{-1}$. 

The source position, with a 1$\sigma$  error of 8\arcsec, lies 6\arcsec{}  from the galaxy \srcnamemlong (hereafter \srcnamemns) and 18\arcsec{}
 from the 11{th} magnitude A star, HD~23235465 (Fig.~\ref{fig:swiftimage}). On September $12$, 2016, \xmm made a pointed
observation of the source and found an X-ray source consistent with the
coordinates of the nucleus of \srcnamem (also known as LEDA~2725953) and no emission from the
star.

It was possible to perform a very crude analysis on the 11 photons in the 
slew spectrum to investigate the gross spectral properties of the 
detection. Detector matrices were calculated, taking into account 
the transit of the
source across the detector, using a technique
outlined in \citet{Read08}. A simple absorbed power-law fit gives a slope
$\Gamma=2.7\pm{1.0}$ assuming no intrinsic
absorption above the Galactic value 
of $1.9\times10^{20}$cm$^{-2}$ \citep{Willingale}.
This corresponds to an absorbed flux of 
$F_{0.2-2}\sim2\pm{0.7}\times10^{-12}$ \fluxUnits using the above model.
We calculate a 2$\sigma$ upper limit from the \rosat All-Sky Survey 
(RASS) 
 at this position of 0.010 count s$^{-1}$ \citep[see][for a description of the upper limit
calculation]{Esquej07}, a factor of 20 lower 0.2--2 keV flux using the same spectral model.

\section{X-ray and UV observations}
\begin{figure}
\centering
\rotatebox{0}{\includegraphics[height=6.5cm]{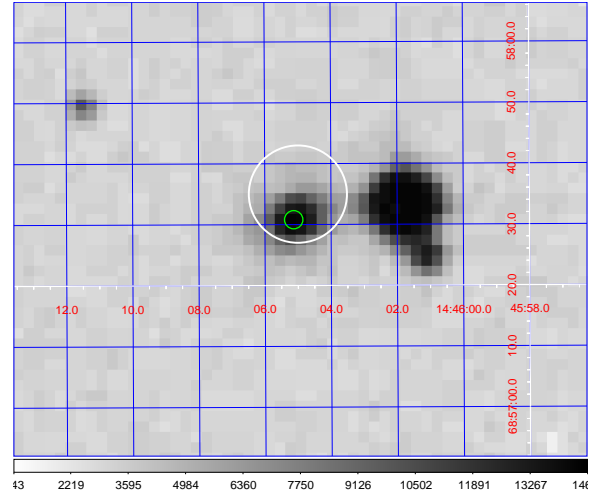}}
\caption[dss image]
{ \label{fig:swiftimage} Digital sky survey image. 
 Shown is  the \xmm slew error circle ($8\arcsec$ radius; white) 
and UVOT-enhanced \swiftns, 90\% confidence, error circle 
($1.5\arcsec$ radius; green) centred on the detections and coincident
with the nucleus of the galaxy \srcnamemlongns. The bright 
object immediately to the west of the galaxy is the 11th magnitude star 
HD~23235465.
}
\end{figure}

An X-ray monitoring programme was initiated with the \ngswift \citep[hereafter \swiftns;][]{Gehrels} to follow the evolution of the source flux and spectrum. Snapshot 3ks observations
were made, initially once a week and then less frequently, with the X-ray 
telescope \citep[XRT;][]{Burrows05} in photon
counting mode and the UV optical telescope \citep[UVOT;][]{Roming}.
 The \swiftns-XRT observations were analysed following the 
procedure outlined in \citet{Evans} and the UVOT data were 
reduced as described in \citet{Poole}. 
An accurate position for the source in the \swiftns-XRT field can be determined
by matching the UVOT field of view with the USNO-B1 catalogue and registering
the XRT field accordingly \citep{Goad}. The resulting source position, 
$\alpha_{J2000}$=14:46:05.13, $\delta_{J2000}$=68:57:30.8 
 ($\pm 1.5\arcsec$; 90\% confidence)
is coincident with the 2MASS position of the galactic nucleus (see Fig.~\ref{fig:swiftimage})
to within the 90\% confidence uncertainty.

In parallel,  five \xmm pointed observations were triggered
between September 8, 2016 and April 25 2018 (observation ID=0763640201, 0763640401, 0763640501, 0763640601, 0823330101).
A summary of observations and exposure times is given in Table~\ref{tab:xobs}.
In each observation, the EPIC-pn and MOS-1 cameras 
were operated in full frame mode with the {\em thin1} filter in place,
while the MOS-2 camera was used in small window mode with the medium 
filter. The source was too
faint for statistically significant data to be collected from the
reflection grating spectrometers. 
 
The XMM data were analysed with the \xmm Science Analysis System 
\citep[SAS v16.0.1;][]{Gabriel}. Light curves were extracted from the 
observations and searched for periods of high background flaring revealing
that the first four observations were considerably affected by background.
Observations 1, 3, and 4 could be cleaned by filtering out periods of high
background, but observation 2 had moderate background for the whole exposure and
has been used in its entirety. For observations 1 and 4 we used a single contiguous
section of data from mission reference time 589743669-589748442 and 596642336-596651995, respectively. For observation 3 we created a light curve over the whole field of view from single-pixel events in the energy range 10-12 keV. We then produced a series of good time intervals where the count rate in this light curve was
$<=1.4$ c/s and used these to extract the source spectrum.

\begin{center}
\begin{table}
{\small
\caption{X-ray observation log of \srcname}
\label{tab:xobs}      
\hfill{}
\begin{tabular}{l c l l}
\hline\hline                 
Mission$^{a}$ & Date & Exp time$^{b}$ & Flux$^{c}$ \\
              &      &   (s)    &    \\
\\
ROSAT-Survey & 1990-07-11 & 958.2 & $<0.07$ \\
XMM-Newton slew & 2004-11-19 & 10.5 & $<0.78$ \\
XMM-Newton slew & 2013-11-03 & 4.9 & $<1.05$ \\
XMM-Newton slew & 2014-04-24 & 8.2 & $<0.99$ \\
XMM-Newton slew & 2014-11-07 & 9.5 & $<0.55$ \\
XMM-Newton slew & 2016-08-22 & 8.0 & $1.68 \pm {0.59}$ \\
XMM-Newton pointed & 2016-09-08 & 3918 & $0.83 \pm 0.03$ \\
Swift                   & 2016-09-12 & 2899 & $1.06 \pm 0.13$ \\
Swift                   & 2016-09-15 & 1071 & $0.50 \pm 0.12$ \\
XMM-Newton pointed & 2016-09-16 & 16140 & $0.84 \pm 0.02$ \\
Swift                   & 2016-09-20 & 2085 & $1.01 \pm 0.20$ \\
Swift                   & 2016-09-27 & 2073 & $1.01 \pm 0.18$ \\
Swift                   & 2016-10-04 & 1731 & $1.31 \pm 0.18$ \\
Swift                   & 2016-10-10 & 1064 & $0.82 \pm 0.16$ \\
Swift                   & 2016-10-13 & 1211 & $0.98 \pm 0.18$ \\
Swift                   & 2016-10-17 & 1963 & $1.13 \pm 0.16$ \\
Swift                   & 2016-10-22 & 879 & $1.05 \pm 0.25$ \\
XMM-Newton pointed & 2016-10-30 & 5990 & $0.95 \pm 0.03$ \\
Swift                   & 2016-11-02 & 2160 & $0.96 \pm 0.13$ \\
Swift                   & 2016-11-23 & 1408 & $0.81 \pm 0.17$ \\
XMM-Newton pointed & 2016-11-27 & 7982 & $0.69 \pm 0.02$ \\
Swift                   & 2016-12-13 & 2138 & $0.49 \pm 0.10$ \\
Swift                   & 2017-01-13 & 682 & $0.21 \pm 0.13$ \\
Swift                   & 2017-01-26 & 1885 & $0.37 \pm 0.23$ \\
Swift                   & 2017-02-17 & 1923 & $0.15 \pm 0.06$ \\
Swift                   & 2017-03-28 & 2210 & $0.126 \pm 0.048$ \\
Swift                   & 2017-04-20 & 1975 & $0.186 \pm 0.057$ \\
Swift                   & 2017-08-14 & 4947 & $0.118 \pm 0.028$ \\
XMM-Newton pointed & 2018-04-25 & 17820 & $0.016 \pm {0.003}$ \\
Swift                   & 2018-12-20 & 2934 & $0.025\pm {0.015} $ \\
\hline                        
\end{tabular}}
\hfill{}
\\
\\
$^{a}$ \xmmns, EPIC-pn camera: slew observations performed in {\em full frame}
mode with
the {\em medium} filter; pointed observations performed in {\em full frame} mode with
the {\em thin1} filter. \swiftns-XRT observations performed in {\em pc} mode. \\
$^{b}$ Useful exposure time after removing times of high background flares. \\
$^{c}$ Absorbed flux in the 0.2-2 keV band, units of $10^{-12}$\fluxUnitsns. 
For simplicity this was calculated using a power-law model of slope 2.5 and 
Galactic absorption of $1.9\times10^{20}$cm$^{-2}$ in all cases \\
\end{table}
\end{center}

\section{Optical observations} 
A set of optical observations were initiated shortly after the 
discovery of \srcnamens.
Photometry was obtained to monitor the flux evolution 
in the optical bands and complement the UVOT observations. Spectroscopic
observations were made to check for the presence of a persistent
AGN and to look for broad or narrow line features which may have been produced
by material reprocessing the nuclear flare emission.

\subsection{Photometry}
\srcname was observed with the ESA Test-Bed Telescope Cebreros  \citep[MPC code Z58;][]{ocana16} with the prime focus 4Kx4K CCD camera using B,V, R Johnson filters.
Short observations of 30--90 seconds  were taken on September 6, 2016, September 20, 2016, and October 31, 2016.
Differential photometry was provided by using up to 2000 nearby stars 
from the catalogue UCAC-4. Photometric residuals are better than 0.1 magnitude.

\subsection{Spectroscopy}

\begin{figure}

\begin{center}



\hspace*{0.0in} \rotatebox{0}{\includegraphics[width=9cm]{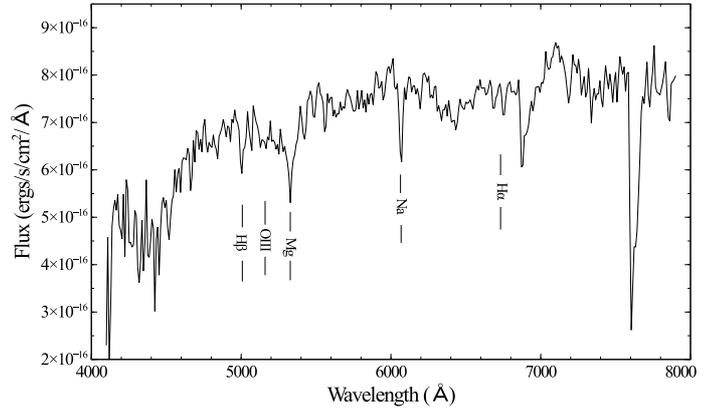}}



\end{center}
\caption[\srcname Liverpool-telescope optical spectrum]
{ \label{fig:LTspec} Optical spectrum (observed frame) of \srcname taken with the 
SPRAT spectrograph on the 2m Liverpool Telescope on September 5, 2016.
}
\end{figure}

A low-resolution ($R\sim350$, $\lambda=4000-8000${\AA}) optical 
spectrum of \srcname was obtained using the  SPRAT spectrograph \citep{sprat}
of the 2.0m Liverpool Telescope \citep{lt} on September 5, 2016. 
The exposure time used was 300 s.  The spectrum was reduced and 
calibrated using the standard pipeline \citep{lt_pipeline}, and showed
a non-active galaxy (Fig.~\ref{fig:LTspec}) at a redshift of z = $0.029\pm{0.001}$.

\begin{figure}
\begin{center}
\hspace*{-0.2in}    \rotatebox{0}{\includegraphics[height=6cm]{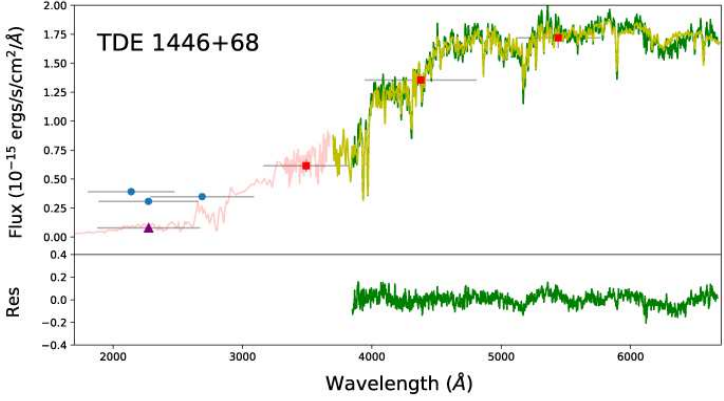}}
  \end{center}
\caption[\srcname FAST/FLWO optical spectra]
{ \label{fig:FLWOspec} Optical spectrum of \srcname 
taken on September 4, 2016 with the FAST instrument on the FLWO 
telescope (light green)
fitted with a stellar population
model shown in dark green and pink (see text for details of the fitting procedure).
The mean \swift UVOT filter fluxes from the period September 12, 2016 to 
August 14, 2017 are 
overplotted as red squares (U,B,V) and blue circles (UVW1,UVM2,UVW2). 
The GALEX-NUV filter point from 2007 is shown as a purple triangle.
}
\end{figure}

\srcname was also observed with the FAST spectrograph on the 60-inch telescope at Fred Lawrence Whipple Observatory (FLWO) on September 4, 2016 (Fig.~\ref{fig:FLWOspec}). The two 900s spectra were extracted and wavelength-calibrated using an automated pipeline, and flux-calibrated using a standard star taken on the same night. These observations confirmed the redshift of $z=0.029\pm{0.001}$.

The stellar population of \srcname was analysed using the
STARLIGHT spectral population synthesis code \citep{Fernand05,Fernand11}. 
The code uses the \citet{Bruzual} single stellar
population models and requires the data to be corrected for foreground
Galactic extinction and taken to the rest frame. We used the \citet{Cardelli} 
extinction curve to correct for foreground reddening and
adopted $A_{V}=0.064$ and shifted the spectrum to the rest wavelength
using $z=0.029$. Figure~\ref{fig:FLWOspec} presents the results from the spectral population synthesis analysis showing the observed and total synthetic
spectra and its residuals. The mean \swift UVOT filter fluxes from the flat part of the UV light curve (September 12, 2016 to August 14, 2017) are overplotted on   the 
synthetic spectrum, as is a GALEX-NUV pre-flare flux from 2007 (see Section 6.1).
The fit needs a relatively young stellar population of $1-5$ Gyr to model the blue flux in conjunction with an older
population. The total initial mass of the galaxy is $1.1\times10^{10}$\msolar and the current mass $5.7\times10^{9}$\msolar after correcting for the gas restored to the ISM.
The residuals clearly show that no emission lines are present in
\srcnamens.


\section{Radio observations}

We observed \srcname at 6.0 GHz and 21.7 GHz with the 
National Science Foundation Karl G. Jansky Very Large Array (VLA) on 2016 September 15.16 UT and 2017 February 22.52 UT. 
The first observation was taken while the VLA was moving from B configuration into its most extended A configuration and used 3C48 as the flux calibrator and 
J1436+6336 as the gain calibrator at both frequencies. The second was taken in the most compact D configuration of the VLA  and used J1438+6211 as the gain calibrator at 6 GHz, J1436+6336 as the gain calibrator at 21.7 GHz, and 3C286 as the flux calibrator at both frequencies. We reduced and imaged the data using standard Common Astronomy Software Applications (CASA) routines \citep{McMullin07}. 

\begin{figure}
 \begin{center}
    \rotatebox{0}{\includegraphics[width=9cm]{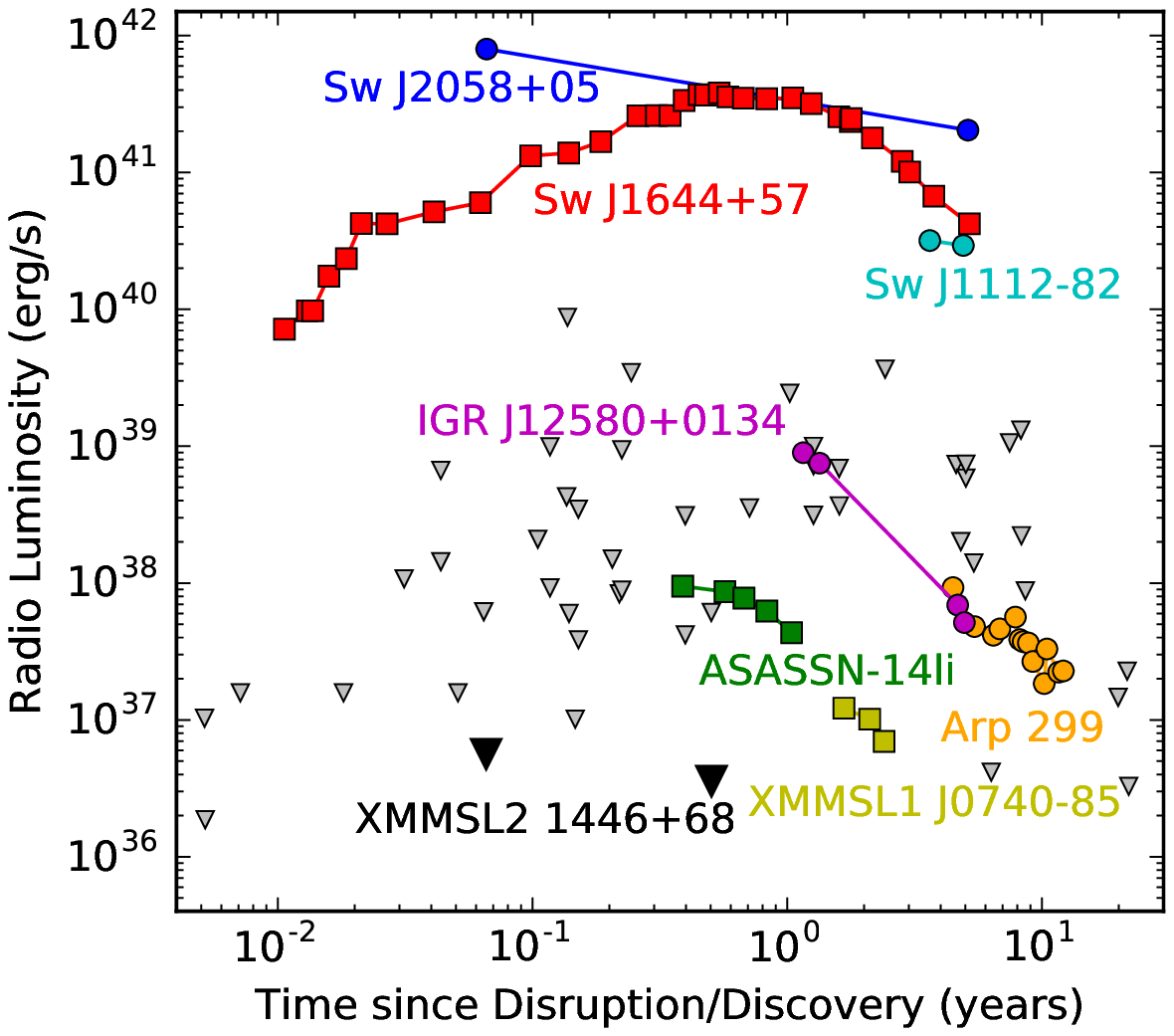}}
  \end{center}
\caption[\srcname The radio luminosity of \srcname compared with other TDEs]
{ \label{fig:radio} Upper limits on the radio luminosity of \srcname (black triangles) compared with other TDEs and plotted against observed time since disruption or discovery. The coloured points are radio detections of TDEs from the literature \citep{Zauderer11, Berger12, Zauderer13, Eftekhari18, Cenko12a, Pasham15, Brown15, Irwin15,Perlman17,Alexander16,Alexander17,Mattila18}, while the grey triangles are upper limits from non-detections \citep[][{Alexander et al. 2019 and references therein}]{Bower13,vanVelzen13,vanVelzen18, Arcavi14, Chornock14, Blagorodnova17, Blanchard17}. All upper limits are $5\sigma$. 
}
\end{figure}

We did not detect any radio emission at the enhanced Swift position of \srcname or at the position of the nearby star HD 23235465. We determined the image root mean square  (rms) noise at the source position using the {\tt imtool} package \citep[part of {\tt pwkit};][]{Williams17}. In the first observation, we find rms values of 9 $\mu$Jy at 6.0 GHz and 28 $\mu$Jy at 21.7 GHz; in the second the rms values are 6 $\mu$Jy at 6.0 GHz and 28 $\mu$Jy at 21.7 GHz. The more constraining $5\sigma$ upper limits on the 6 GHz radio luminosity of \srcname are shown in Fig.~\ref{fig:radio} in comparison to other TDEs. The radio limits for \srcname are the deepest radio limits for any X-ray TDE obtained to date, and the deepest limits for any TDE on timescales of a few months. Only one optical TDE, iPTF16fnl, was found to have  a deeper radio limit just days after discovery \citep{Blagorodnova17}. However, iPTF16fnl remains the faintest and fastest-fading optical TDE discovered to date, and therefore may not be representative of the typical population. \srcname is $\gtrsim10^{5}$ times fainter than the radio-loud jetted TDE \swtd and $\gtrsim20$ times fainter than the canonical radio-weak TDE ASASSN-14li at comparable epochs \citep{Berger12,Alexander16}, making it highly unlikely that \srcname launched either a relativistic jet or a sub-relativistic outflow.

\section{X-ray light curve}
\label{sec:var}
\begin{figure}
 \begin{center}
    \rotatebox{0}{\includegraphics[width=9cm]{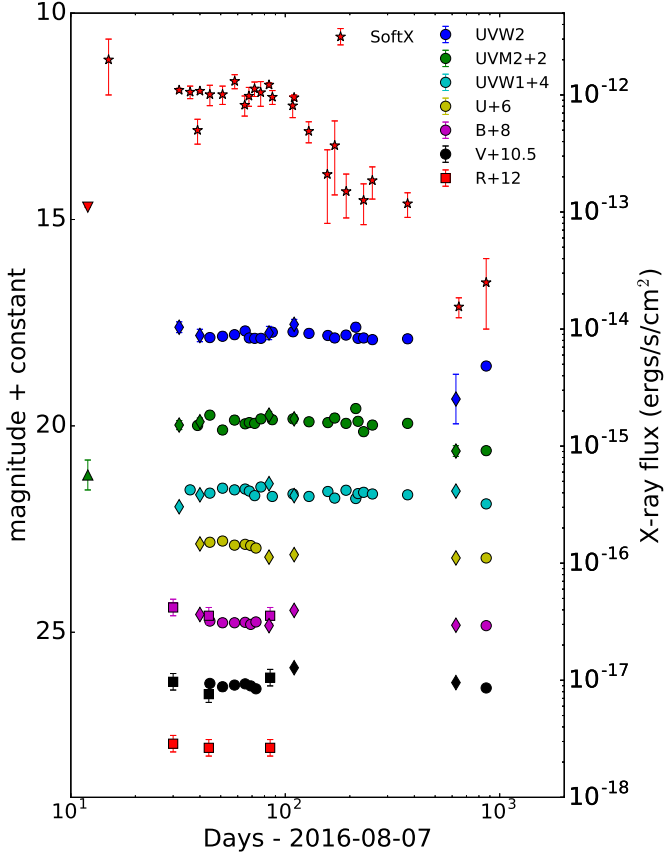}}
  \end{center}
\caption[\srcname long-term light curve]
{ \label{fig:lcurve} Long-term X-ray, UV, and optical light curve 
of \srcname plotted against time. From top to bottom, 0.2-2 keV, UVW2 (1928\AA), UVM2 (2246\AA), UVW1 (2600\AA), U (3465\AA), 
B (4500\AA), V (5430\AA), and R (6580\AA). X-ray points are shown as 
red stars, \swift-UVOT filters as   circles, \xmm-OM
filter points as diamonds, and the Cebreros telescope points as squares.
A GALEX-NUV measurement, taken in 2007 and calibrated to the UVM2 filter
values, is shown as a green triangle and the RASS soft X-ray upper limit from 1990-1991 
as a downward red triangle. All points show the observed fluxes and magnitudes without host galaxy subtraction.
}
\end{figure}

In Fig.~\ref{fig:lcurve} we show the historical  
light curve of \srcnamens. The soft X-ray (0.2-2 keV)  flux is effectively constant for
the first $\sim90$ days, with a temporary drop of a factor 2 after 24 days, which we
attribute to small variations in the local accretion rate. It then drops by a factor of 10 over the following $\sim100$ days and a further factor of 10 by day 600. Modelling the observed X-ray emission as a constant, from the
discovery date $t_{0}$=August 22, 2016 to a later date $t_{drop}$ when the flux 
follows a power-law decay, we find that at   time $t$

\begin{equation}
  f(x)=\begin{cases}
    1.06\pm{0.03}\times10^{-12} & \text{if $t_{0}<t<t_{drop}$},\\
    9.3^{+10.6}_{-5.0}\times10^{-9}  \left( t - t_{0} \right)^{-2.02\pm0.16} & \text{if $t\ge t_{drop}$},
  \end{cases}
\end{equation}
where $(t-t_{0})$ is in units of days and $t_{drop}$ = November 19, 2017$\pm{\text{2 days}}$.

\begin{figure}
  \begin{center}

\begin{minipage}{3in}
    \rotatebox{-90}{\includegraphics[width=6cm,trim=0.0cm 1.0cm 0.0cm 0.0cm,clip]{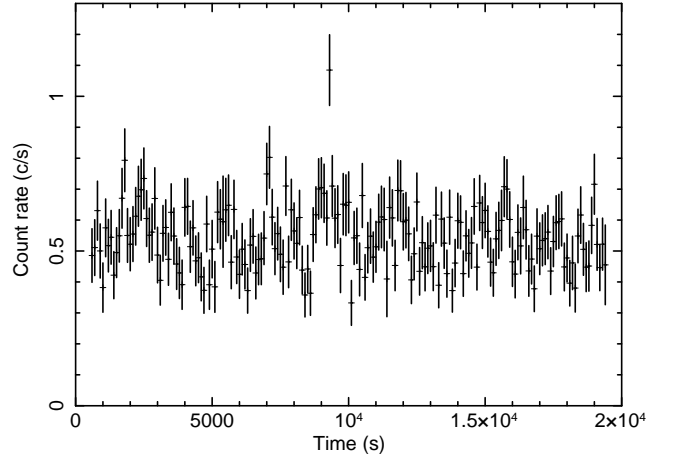}}
\end{minipage} 

  \end{center}
\caption[\srcname Short term light curve]
{ \label{fig:lcurve_short} Background subtracted, exposure corrected, 
EPIC-pn, 0.2--2 keV light curve for the September 16, 2016  
\xmm pointed observation, displayed in 100s bins.}
\end{figure}

In Fig.~\ref{fig:lcurve_short} we show the exposure-corrected, background-subtracted,
0.2-2.0 keV short-term light curve from the \xmm pointed observation of 
September 16, 2016. 
The X-ray flux shows small variations on short timescales, which can be used to obtain
an estimate of the black hole mass.
We   used the method of \citet{Ponti} on a 10 ks segment of the September 9, 2016 \xmm
pointed observation, finding an insignificant excess variance of $0.007\pm{0.02}$, 
from which we extracted a 90\% confidence lower limit on
the mass of $M_{BH}\ge2\times10^{6}$\msolarns. We note, however, that this technique was calibrated on AGNs, which may not be strictly applicable to the variability seen in a TDE.

Quasi-periodic oscillations (QPO) have been found  in the X-ray light curves of some TDEs \citep{Reis12,Pasham19,Lin13}. We barycentred the event file of the \xmm observation of September 9, 2016, created source and background time series, and searched for a QPO in the power spectrum. No significant peaks were found.  

\subsection{UV light curve}
\label{sec:uv}

During the five pointed \xmm observations, the optical monitor (OM)
cycled between the {\it B}, {\it U}, {\it UVW1}, {\it UVM2}, and {\it UVW2} filters.
\swiftns-UVOT observations were 
performed with the {\it UVW1, UVM2}, and {\it UVW2} filters, except for the first six observations
and the final observation, which also used the optical U,B,V  filters. 
The galaxy was detected in all filters in all the observations except for the
final \xmm observation, which did not detect the galaxy in the {\it UVW2} filter
using the source search software. A faint object is visible by eye, however, and so
the magnitude of this source was extracted by analysing the image directly.
Relative filter fluxes were determined using several nearby sources of
comparable brightness as references. The absolute flux scale was taken from 
the \swiftns-UVOT filters, with the \xmmns-OM points scaled to these values.

\begin{figure}[ht]
\begin{center}
\hbox{
\rotatebox{-90}{\includegraphics[height=3.cm]{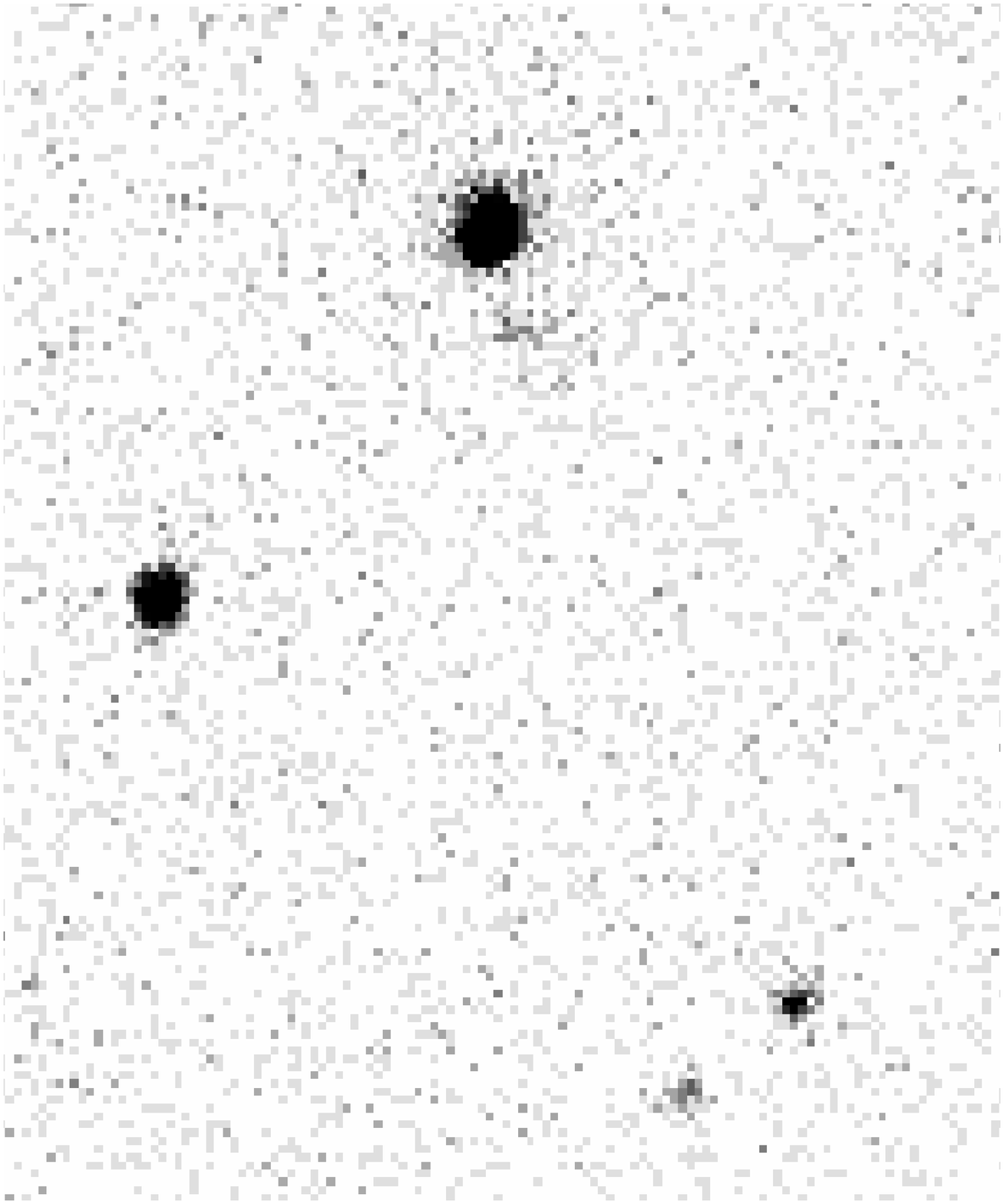}}
\hspace{0.1cm}
\rotatebox{-90}{\includegraphics[height=3.cm]{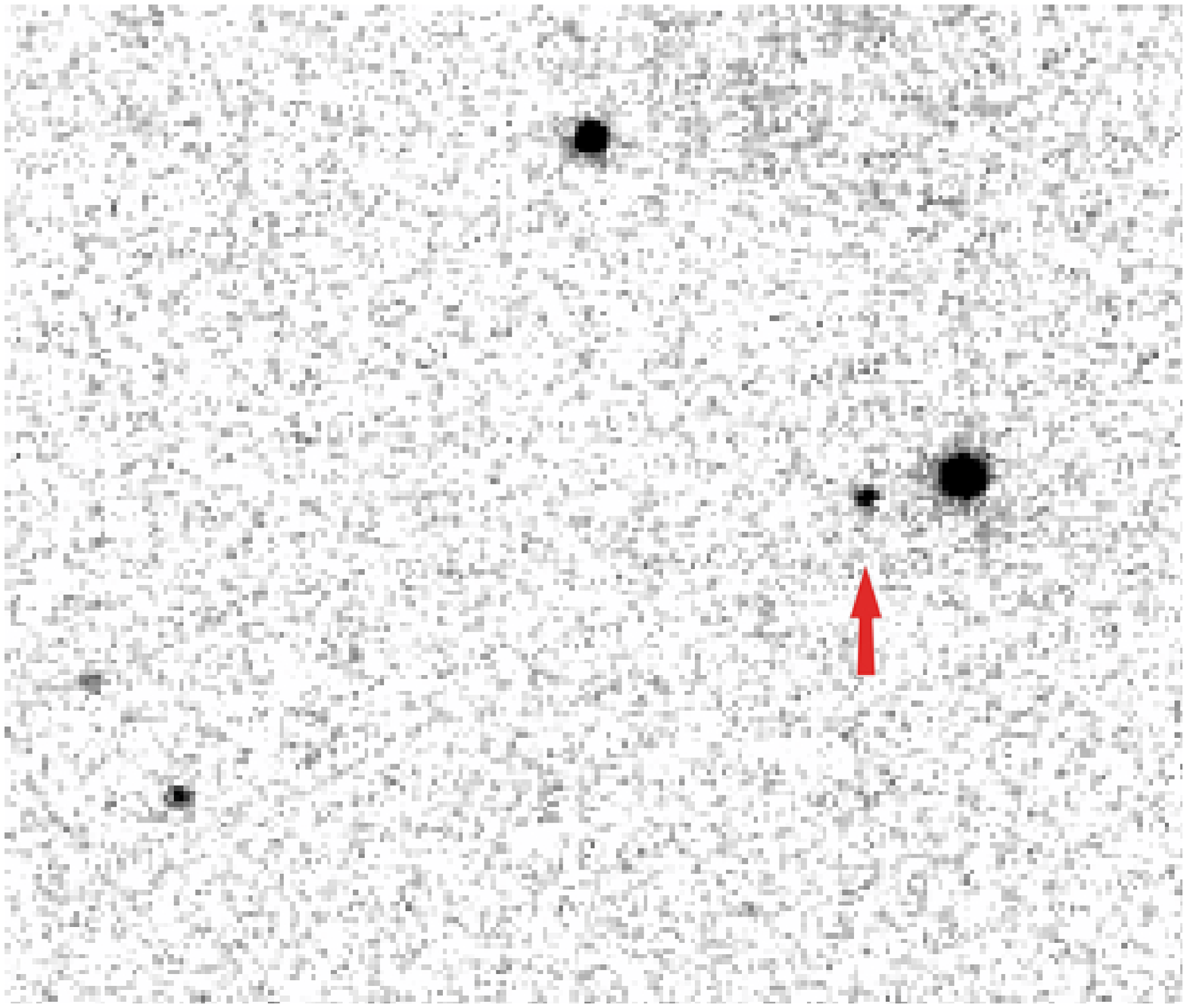}}
\hspace{0.1cm}
\rotatebox{-90}{\includegraphics[height=3.cm]{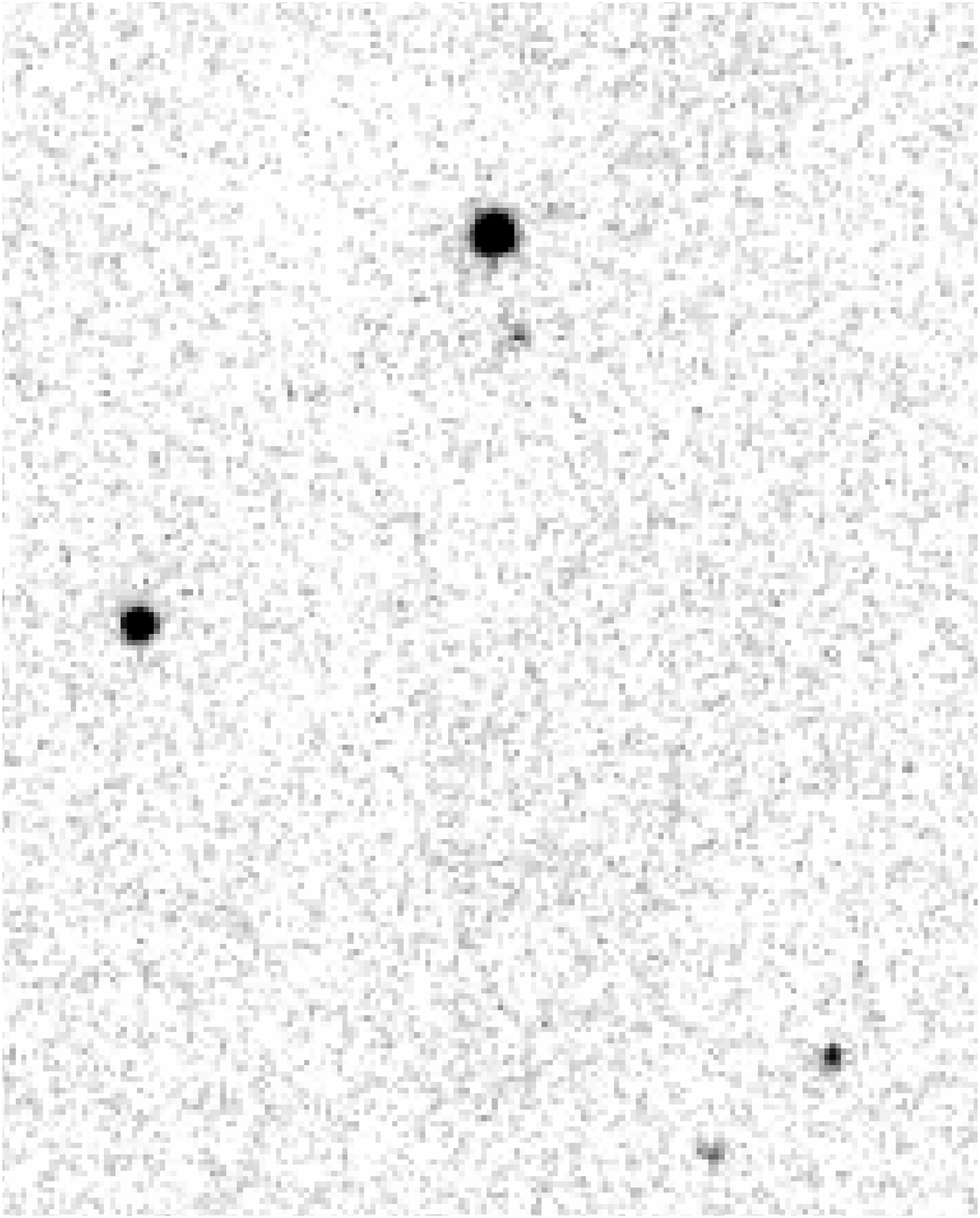}}
}
\end{center}
\caption[Near-UV images of \srcname]{
GALEX-NUV image taken in 2007 (left), XMM-OM UVM2 image from September 30, 2016 (middle; \srcname 
indicated by red arrow), and
XMM-OM UVM2 image from Aptil 25, 2018 (right).
}

\label{fig:uvimages}
\end{figure}

GALEX observed the position of the galaxy between 2005-03-05 and 2007-02-18.
There is no detection of the galaxy present in the catalogue \citep{Bianchi}; 
however, a faint enhancement is visible by eye 
in the near-UV (NUV filter) image (2267 \AA)  lying within the radial 
profile of the bright A star, HD~23235465.
By using the catalogue magnitudes of the surrounding stars and galaxies, and 
an analysis of the count rates in this image we calculate that the NUV 
magnitude 
of the galaxy was $21.0\pm{0.15}$ (AB magnitude) during the GALEX observations,  which is  
$1.2$ magnitudes below the OM-UVM2 magnitude from September 8, 2016. The NUV filter has a similar bandpass to the UVM2 filter and gives consistent magnitudes to within $\Delta m=0.026$, with a conservative standard deviation due to 
colour-dependence and other systematic effects of 0.33 magnitudes \citep{pageuvm}.
To illustrate the variation in the UVM2/GALEX-NUV flux, in Fig.~\ref{fig:uvimages} we
show images of the field around \srcname in the GALEX observation and the OM-UVM2
observations from September 30, 2016 and April 25, 2018. 

In Fig.~\ref{fig:lcurve} we see that the UVW2 and UVM2 fluxes remained roughly
constant for the first $\sim 400$ days, but had fallen by $\sim1$ magnitude by day 600.\footnote{We confirmed that none of the Swift-UVOT observations fell on the regions
of low sensitivity reported in \citet{Edelson15}.}
Using the GALEX-NUV observation as a measurement of the quiescent galaxy flux we find 
a galaxy-subtracted UVM2 flux of $1.2\times10^{-13}$ \fluxunits.
The optical filters do not vary significantly over the 850 days of monitoring, indicating that the optical flux is dominated by the host galaxy. 

\subsection{X-ray spectral analysis}
\label{sec:specfit}
We extracted spectra from the second \xmm pointed observation (hereafter XMM2),
which was taken 25 days after discovery and has the highest exposure time, from circles with optimum radii determined by the task {\em eregionanalyse}
for the EPIC-pn, MOS-1, and MOS-2 cameras.
Background spectra were created from nearby source-free regions.
The source spectra were grouped to have a minimum of 20 counts
per bin and were fitted simultaneously within the XSPEC
package (v12.8.2). 
Fits were performed using the chi-squared statistic over the energy range  
0.3--10 keV. Quoted errors
are at the  90\% confidence level unless otherwise stated.
 
As a first step we fit the XMM2 observation with a simple power-law model 
and galactic absorption of $1.9\times10^{20}$cm$^{-2}$, modelled by TBABS with elemental abundances set to those in \citet{Wilms}.
The fit is good (\chired=354/347) with a slope of $\Gamma=2.59\pm{0.03}$.
If we fit only over the 2--10 keV range, the slope flattens to $2.48\pm{0.26}$
and small residuals are present at lower energies.
We attempted to model the low-energy excess as an extra emission component,
with a black body of $kT=132^{+20}_{-14}$eV (\chired=344/345).
The extra component is not statistically significant. We also tried modelling the
soft component with a Bremsstrahlung (\chired=347/345; $kT=428^{+63}_{-75}$
eV) and power law (\chired=354/345).
We repeated this for the five \xmm pointed observations\footnote{Only observations 1, 2, 3, and 4 had sufficient statistics to warrant the inclusion of the
MOS camera data. Observation 5 was fit with the EPIC-pn data alone.}, finding that an absorbed
power law provided a satisfactory fit for each. In Table~\ref{tab:specfits} we show
the results of these fits, all of which are consistent with a power-law slope of 2.5--2.6. 

\begin{figure}
  \begin{center}
    \rotatebox{0}{\includegraphics[width=8cm]{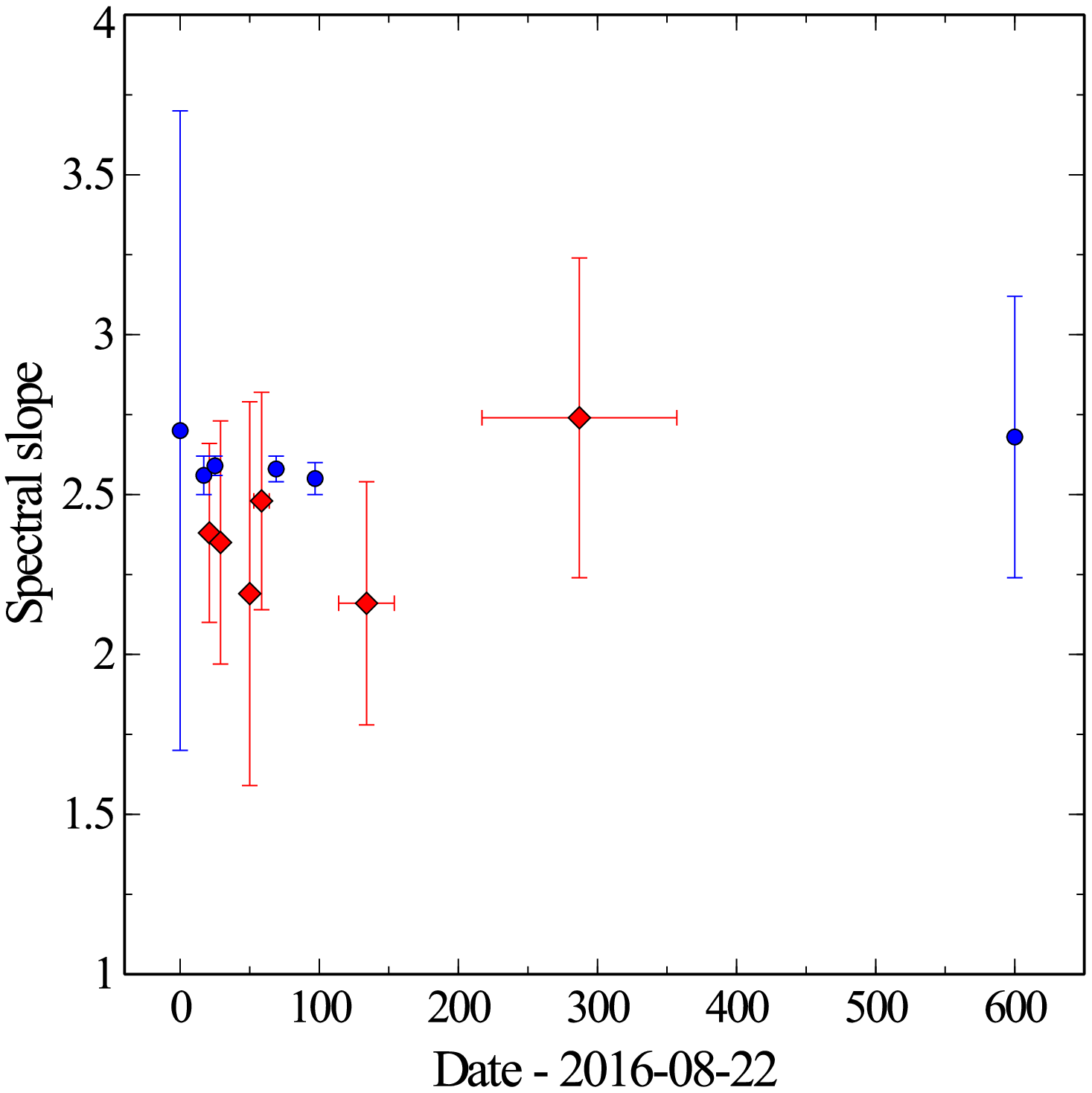}}
    \rotatebox{-90}{\includegraphics[width=6cm]{hardratio_timev2.ps}}
  \end{center}
  \vspace{0.5cm}
\caption[\srcname X-ray spectral slope]
{ \label{fig:slopes_hratio} Upper: 0.3--10.0 keV spectral slope from absorbed power-law fits to the \xmm (blue circles) and \swift (red diamonds) observations of \srcnamens. Some of the \swift observations have been combined to reduce the error bar size. Lower: Source spectral hardness expressed as the ratio of 
the 1.5-8 keV to 0.3-1.5 keV fluxes.}
\end{figure}

The same fitting process was applied to the \swift XRT observations from 0.3--10 keV.
Figure~\ref{fig:slopes_hratio} shows the fitted slope of each observation, which is effectively
constant, within the errors, between discovery and 600 days later.

\begin{figure}
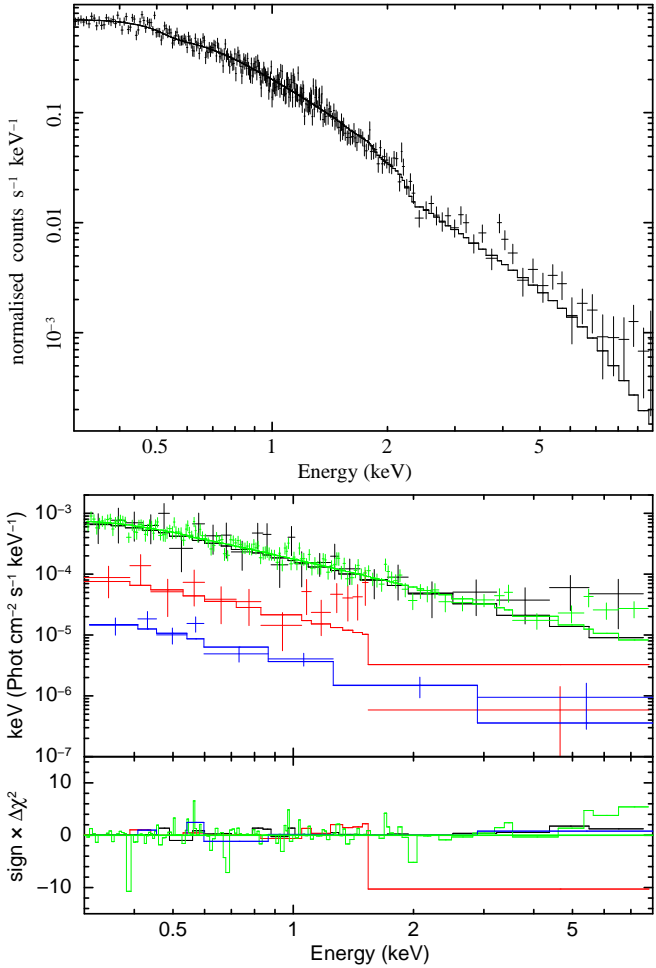

\centering
\rotatebox{-90}{\includegraphics[height=9cm]{spec_xmm_comb.ps}}
\vspace{0.5cm}
\rotatebox{-90}{\includegraphics[height=9cm]{multispectra.ps}}
\vspace{0.5cm}
\caption[A fit of the combined EPIC-pn spectrum to a power-law.]
{ \label{fig:xspec} Upper: Fit of the EPIC-pn spectrum combined from the 
first four \xmm observations to a power law absorbed by the Galactic column
of $1.9\times10^{20}$ cm$^{-2}$. Lower: Spectra from the first \swift
observation (September 12, 2016; black), the \xmm observations of 
November 27, 2016 (green)
and April 25, 2018 (blue), and the combined data from \swift between February 15, 2017 and August 14, 2017 (red), all fitted with a power law of $\Gamma=2.58$
absorbed by the Galactic column. Residuals to the fits are shown below.}
\end{figure}

To improve the statistics we combined the counts from the clean data in observations 1, 3, and 4 from the \xmm EPIC-pn camera. This yielded a 
power-law slope of $2.58\pm{0.04}$ and $\chi^{2}=165/183$.
Using the full EPIC-pn dataset from observations 1-4, without applying any background filtering, we get a consistent slope of $\Gamma=2.58\pm{0.03}$
and $\chi^{2}=425/410$ suggesting that the periods of higher background do not cause significant spectral deviations (Fig.~\ref{fig:xspec}). We then used this `dirty' combined spectrum of 45.9 ks to look for spectral features beyond the simple power law. 
A small improvement can be found by adding a black-body component ($\chi^{2}=414/408$) with $kT=113^{+12}_{-13}$ eV when the power-law 
slope reduces to $2.43\pm{0.08}$. In this case $13\pm{6}$\% of the 
total 0.2--2 keV flux is provided by the soft
component. A Comptonised black-body model \citep[{\tt COMPBB} in {\tt XSPEC}; ][]{compbb} gives a similar 
result ($\chi^{2}=416/408$) with a very low black-body temperature 
($kT=10^{+10}_{-8}$ eV), electron temperature ($kT_{e}=35^{+4}_{-7}$ keV), and
optical depth ($\tau=0.31^{+0.24}_{-0.17}$).
In summary, the X-ray spectrum is dominated by a somewhat steep power law 
of slope $\sim2.6$ with weak evidence for a faint excess at low energies.
The fifth \xmm observation has low statistics, but gives a consistent 
spectral slope of $\Gamma=2.68^{+0.48}_{-0.40}$, to within the errors. 

The unabsorbed 0.2--10 keV luminosity from the \xmm slew observation of August 22, 2016 
fit with a power law of $\Gamma=2.58$ is $L_{X}=6\times10^{42}$\lumunits.
The bolometric correction is quite uncertain for this spectrum and we need to model the
full dataset carefully to try and estimate the flux that emerges in the EUV band.
To find the UV flux from the flare we subtract the contribution of the stellar population of the host galaxy using the GALEX flux for the UVM2 filter (see section~\ref{sec:uv}),
and use the fifth \xmm observation for the UVW2 filter. The contribution of the galaxy
to the total filter flux, from the first \xmm observation, is then 
20\% (UVW2) and 33\% (UVM2).
Under the assumption that the UV and X-ray emission both come from a coherent
accretion structure, we fit the galaxy-subtracted UVM2 and UVW2 data and the
X-ray spectrum from XMM observation 1, with the optxagnf model \citep{Done12}. 
This model consists of a multi-colour disc and a corona of optically thick 
electrons which Compton scatter disc photons to X-ray energies\footnote{It also includes
a power-law component, which we set to zero in the fits}. 
Given the number of free parameters in the
model we assume a non-spinning black hole ($a=0$) and use a prior for the black hole mass. 

From the relationship of black hole mass to bulge K-band luminosity
\citep{MarconiHunt} we get $M_{BH}\sim5\times10^{6}M_{\odot}$ from the point source
2MASS magnitude of $m_{K}=13.06\pm{0.05}$ (or 12.2 extended), with a systematic 
error of 0.3 dex ($2\times10^{6}M_{\odot}$). Using the current galaxy stellar mass of
$5.7\times10^{9}$\msolar, we find a mass of $M_{BH}=2\times10^{7}M_{\odot}$ \citep[rms error of 0.6 dex;][]{ReinesVolonteri}. The black hole mass in \srcnamem has 
also been calculated from the $M_{BH}-\sigma$ relation \citep{Wevers19} as $M_{BH}=7^{+17}_{-5}\times10^{7}M_{\odot}$. We use the lower and higher of these estimated masses in the optxagnf fits 
(see table~\ref{tab:optxagn}). The bolometric luminosity from the XMM1 observation 
can be obtained from these fits
using $L_{bol}=1.3\times10^{38}M_{BH}\dot{m}$, where $\dot{m}$ is the accretion
rate in units of the Eddington accretion rate, giving $L_{bol}=1.0(1.3)\times10^{43}$
\lumunits for a black hole mass of $5\times10^{6}$ ($7\times10^{7}$) \msolar.
The radius where emission transitions from a thermal black-body spectrum to a Comptonised spectrum is 
46.3 (100) $R_{g}$ for the fits. At this radius, very little of the thermal emission falls
within the soft X-ray band, which is therefore dominated by Compton  
upscattering of the UV photons. Using the same model, the Swift observation of
August 14, 2017 (which has a similar UV flux to XMM1, but an X-ray flux lower by a factor of 10)
has $L_{bol}\sim30\%$ that of XMM1, with a lower optical depth of the Comptonising electrons. The absorption-corrected luminosity in the UVM2 and UVW2 filters, after subtraction of 
the quiescent galaxy flux, was  $L_{uv}=2-3\times10^{41}$\lumunits in both observations.

As a further test of the stability of the X-ray spectrum we calculated the ratio between a soft (0.3--1.5 keV) and hard (1.5--8 keV) energy band for all of the X-ray observations (Fig.~\ref{fig:slopes_hratio}).
In this plot we see an apparent strong softening of two spectra produced from the
combination of \swift observations 00034707017, 00034707018, 00034707019 
(February 15, 2017 to March 13, 2017) and 
00034707020, 00034707021, 00034707022 (March 26, 2017 to August 14, 2017). 
The data from \swift observations 
00034707017-00034707022 were combined into a single spectrum
 to maximise the signal-to-noise ratio, and compared with the
last two \xmm observations and the first \swift observation, all fit with
an absorbed power law of $\Gamma=2.58$ (Fig.~\ref{fig:xspec}).
The combined \swift spectrum does not appear to be softer than the other spectra, but rather to
be lacking  flux above 1.5 keV. It may be fit with a power law ($\Gamma=2.69\pm{0.44}$; $\chi^{2}=46/39$) or alternatively by a black body (kT=$225^{+45}_{-33}$ eV;  
 $\chi^{2}=43/39$), both absorbed by the Galactic column.  
 Adding an empirical edge of energy ($E=1.62^{+0.37}_{-0.10}$ keV) and optical depth ($\tau=10.0$) to a $\Gamma=2.58$ power-law model improves the fit (to $\chi^{2}=37/38$).
We note that this trend is {\em \emph{not}} what is expected from cold absorption,
which would have exactly the opposite effect on an X-ray spectrum.
Detailed photo-ionisation modelling of
the TDE in NGC 5905 has shown that individual deep absorption edges of high column density, ionised
material can potentially imprint sharp spectral features \citep[e.g.][]{KomossaBade}.
However, the deepest edge is that expected from oxygen, around 0.7 keV rest frame
\citep[see also][]{Kara18}.
While we formally can fit the spectra with a single absorption edge at 1.62 keV,
this does not correspond to any known transition, and we consider this solution highly artificial,
and do not discuss it further.  Alternative interpretations could involve changes in the property
of the accretion disc corona, which would primarily affect the hard X-ray photons. However,
at present we do not have enough data to constrain this scenario further.
 
\renewcommand{\arraystretch}{1.5}
\begin{table*}[ht]
{\small
\caption{Spectral fits to \xmm pointed observations of \srcnamens.}
\label{tab:specfits}      
\begin{center}
\begin{small}
\begin{tabular}{cccccc}
\hline\hline                 
\\
Date & $\Gamma^{a}$ & Norm$^{b}$ & Flux (0.2-2 keV) & Flux (2-10 keV) & $\chi^{2}$/dof \\
         & & keV$^{-1}$ cm$^{-2}$ s$^{-1}$ & $10^{-13}$\fluxUnits & $10^{-13}$\fluxUnits & \\
\hline\noalign{\smallskip}

2016-09-08 & $2.56\pm{0.06}$ &  $2.27\pm(0.10)\times10^{-4}$ & $8.4\pm{0.3}$ & $2.6\pm{0.3}$ & 116/99\\
2016-09-16 & $2.59\pm{0.03}$ &  $2.29\pm(0.07)\times10^{-4}$ & $8.6\pm{0.2}$ & $2.5\pm{0.2}$ &354/347 \\
2016-10-30 & $2.58\pm{0.04}$ &  $2.52\pm(0.08)\times10^{-4}$ & $9.7\pm{0.2}$ & $2.8\pm{0.2}$ & 241/237\\
2016-11-27 & $2.55\pm{0.05}$ &  $1.85\pm(0.06)\times10^{-4}$ & $7.0\pm{0.2}$ & $2.2\pm{0.2}$ & 228/220\\
2018-04-25 & $2.68^{+0.48}_{-0.40}$ &  $4.0\pm(1.0)\times10^{-6}$ & $0.16\pm{0.03}$ & $0.04^{+0.04}_{-0.02}$ & 15/13\\
Combined$^{c}$ & $2.58\pm{0.03}$ &  $2.20\pm(0.03)\times10^{-4}$ & $8.5\pm{0.2}$ & $2.5\pm{0.2}$ & 425/410\\
\noalign{\smallskip}\hline
\end{tabular}
\\
\end{small}
\end{center}
Fit to the broad-band \xmm spectra, from 0.3--10 keV, of a power law absorbed by the Galactic column (model {\tt TBABS}, $N_{H}=1.9\times10^{20}$cm$^{-2}$). Errors are at the 90\% confidence level. Observations of 2016-09-08, 2016-09-16, 2016-10-30, and 2016-11-27 use the EPIC-pn, MOS1, and MOS2 data; the 2018-04-25 observation only uses the EPIC-pn data.\\
$^{a}$ Power-law slope, $^{b}$ power-law normalisation,
$^{c}$ Combination of observations 1-4 (2016-09-08 -- 2016-11-27), EPIC-pn data only, without filtering for periods of higher background. 
}
\end{table*}

\begin{table*}[ht]
{\small
\caption{Spectral fits with the optxagnf model to UV and X-ray data of \srcnamens.}
\label{tab:optxagn}      
\begin{center}
\begin{small}
\begin{tabular}{ccccccc}
\hline\hline                 
\\
Date & $M_{BH}$ & log L/$L_{edd}^{a}$ & $r_{cor}^{b}$ & $\tau^{c}$ & 
$L_{bol}$ & $\chi^{2}$/dof \\
         & (\msolar) & & ($R_{g}$) & & \lumunits & \\
\hline\noalign{\smallskip}

2016-09-08 & $5\times10^{6}$ &  $-1.80^{+0.03}_{-0.06}$ & $46.3\pm{8.0}$ & $1.96^{+0.03}_{-0.04}$  & $1.0\times10^{43}$  & 191/89 \\
2017-08-14 & $5\times10^{6}$ &  $-2.28^{+0.21}_{-0.17}$  & $58^{+42}_{-49}$  & $1.44^{+0.23}_{-0.48}$  & $3.4\times10^{42}$  & 8/15 \\
2016-09-08 & $7\times10^{7}$ &  $-2.83\pm{0.02}$ & $100^{+0}_{-9}$  & $2.29\pm{0.04}$ & $1.3\times10^{43}$ & 215/89 \\
2017-08-14 & $7\times10^{7}$ &  $-3.35^{+0.05}_{-0.20}$  & $100^{+0}_{-65}$  & $1.66^{+0.18}_{-0.23}$  & $4.1\times10^{42}$  & 10/15 \\

\noalign{\smallskip}\hline
\end{tabular}
\\
\end{small}
\end{center}
Fit to the galaxy-subtracted UVM2 and UVW2 filter count rates and the X-ray spectrum from the \xmm observation of 2016-09-08 and the Swift observation of 2017-08-14 with the
{\em optxagnf} \citep{Done12} model. The hard power-law fraction is set to zero, the electron temperature
($kT_{e}$) is fixed at 10 keV, the disc outer radius is fixed at $10^{3} R_{g}$, and
a Schwarzschild non-spinning black hole is assumed. 
$^{a}$ Log of the accretion rate in units of the Eddington limit.
$^{b}$ The disc radius at which the transition from colour-corrected black-body emission
to a Comptonised spectrum occurs in units of the gravitational radius ($GM/c^{2}$).
The model applies a hard cut-off at 100 $R_{g}$.
$^{c}$ The optical depth of the 10 keV electrons.
}
\end{table*}
\renewcommand{\arraystretch}{1.0}

\section{Discussion}

The lack of emission lines in optical spectra
makes it unlikely that \srcname was a steadily emitting AGN before the
2016 flare. 
The Wide-field Infrared Survey Explorer (WISE) colours, W1 - W2 = 0.0, are also suggestive of a non-active
galaxy \citep{Stern12}, and with a peak bolometric luminosity of $\sim10^{43}$
\lumunits it seems more likely that this event was caused by a TDE. 
From the density of WISE objects we calculate a 0.6\% probability that there
is a foreground object lying within 1.5" of the galaxy nucleus. The possibility of galactic
sources mimicking a TDE spectrum and light curve was discussed in the early TDE literature 
\citep[e.g.][]{KomossaBade,Komossa99b} and the only remaining credible galactic TDE impostor is a nova \citep[e.g.][]{Mainetti16}. The rate of galactic novae is estimated to be 50 per year \citep{Shafter17}, and as the space density of galaxies is 0.0177 Mpc$^{-3}$
\citep{Driver05} then the
probability of one of these novae lying within 1.5'' of the nucleus of a galaxy out to a distance of 127 Mpc is $3\times10^{-5}$ per year. As the slew survey covers $\sim10\%$ of the sky each year and has been running for ten years, then
the probability that the slew survey has seen one such event is $3\times10^{-5}$. Therefore, we conclude that the flare most likely originates within the galaxy \srcnamemns. 

If we integrate the luminosity over the full light curve we obtain a total emitted bolometric luminosity of $4\times10^{50}$ ergs. As we have no knowledge of the flux prior to discovery, this is necessarily a lower limit.
This total energy release is equivalent to a
consumed mass of $2\times10^{-3}$\msolar assuming a conversion efficiency of $\eta=0.1$.
While we may well have discovered the event post-peak and so be underestimating
the total emitted radiation, this result does
agree with the trend seen in many other TDEs where the radiated mass is just a fraction of a percent of a solar mass \citep[e.g. ][]{Komossa1242,Holoien18,Saxton17,Maksym10}. In some earlier work this low
integrated luminosity has led to very low estimates of the mass of the disrupted object, and suggestions that a giant planet or the outer layers of an evolved star 
had been accreted rather than debris from a main-sequence star \citep{Li02,
Walter13}. The apparent ubiquity of the low integrated luminosity \citep[only 3XMM, J150052.0+015452 has so far radiated the equivalent of a large fraction of a
solar mass; ][]{Lin17}  argues instead for a mechanism where the majority of the disrupted
mass is not promptly accreted or where the multi-wavelength radiation is strongly suppressed. Some evidence for this is given by late-time IR emission produced
by the heating of material lying at $\sim1$ pc from the nucleus, which implies
a true emitted luminosity  ten times higher than that seen directly \citep[e.g.][]{Komossa09,vanVelzen16b}.

This particular TDE is unusual in two key ways: a) the X-ray flux begins to fade 
long before the UV flux; b) the X-ray spectrum is characterised by a single steep ($\Gamma\sim2.6$)
power law rather than  a soft pseudo-thermal component or  a hard power law of $\Gamma\sim1.3-2.3,$ as seen in the radio-loud  TDE \citep{Walter13,Burrows11,Cenko12a}.

\citet{Piran15} describe a scenario, backed up by simulations \citep{Shiok15,Bonnerot17},
where the UV and optical emission is produced by shocks from inter-stream collisions,
while the X-ray radiation is produced by the {\em \emph{later}} accretion of this same material.
This model predicts a delay in the X-rays with respect to the UV.
Support was given to this model by the X-ray flux apparently lagging behind the
UV flux by 32 days in ASASSN-14li \citep{Pasham17} and by a delay in the 
onset of observed X-rays in ASASSN-15oi \citep{Gezari17}. 

Another attempt to describe the mutual exclusivity of 
optical and X-ray TDEs is given by the unified TDE model \citep{Dai18}, which
conjectures that X-ray TDEs are 
seen when the viewing angle looks down a funnel directly onto the accretion
structure, while optical TDEs are produced by the 
reprocessing of accretion-driven X-ray photons \citep[see also][]{MetzStone16, Roth16}. In this case the UV should be 
simultaneous with the X-rays, or a little delayed due to light-travel time, as the reprocessing is more or less instantaneous. 

Neither of these models naturally explains the dichotomy between the X-ray and UV
light curves in this TDE. We need a mechanism that maintains the  near-UV luminosity from 
the TDE at an approximately constant value of $\sim3\times10^{41}$\lumunits  for 400 days while the X-ray luminosity
drops by an order of magnitude. As the X-ray spectrum is consistent between the
5 \xmm observations and the \swift observations before February 15, 2017, the
reduction in flux in these observations is likely to be intrinsic rather than due to intervening absorption.
 We note that the combined \swift spectrum of February 15, 2017 to August 14, 2017 (days 177-356), has a relative deficit of photons at high energies, but we are unable to find a physically meaningful interpretation for this spectrum.

If the UV were purely caused by X-ray reprocessing then it should scale with the
X-rays. On the other hand, if the UV is generated from inter-stream shocks then this would indicate a constant rate of return of material over the first 400 days. 

A possible solution could be that the matter around the black hole formed a disc
that spread outwards and accreted viscously \citep{Cannizzo90}. \citet{vanVelzen18} 
saw evidence for this in late-time UV measurements of ten TDEs, consistent with 
a strong flattening of the far-UV and near-UV   emission after 1--2 years and an average 
black-body temperature of $10^{4-5}$K. They showed that this could result in a UV
light curve that decayed very slowly ($t^{-0.1}$) with a roughly constant temperature.
They predicted a low X-ray flux for their TDE sample based purely on the expected emission
from disc thermal processes. We see in \srcname and other events \citep{Saxton12b,Lin17} that Compton upscattering of thermal photons by a warm electron population provides an important contribution to the X-ray emission at all energies that cannot be neglected.
In section~\ref{sec:specfit} we saw that the fading X-rays could
be reproduced by a reduction in the optical depth of a population of Comptonising electrons, while the seed UV photons from the disc remain fairly constant. 

The variety of X-ray spectra that are seen in TDEs is striking. Many events, for example 
NGC 5905, RX J1242.6-1119,  RX J1420.4+5334, NGC 3599, SDSS 1323+48, ASASSN-14li, 
and 3XMM J150052.0+015452, were dominated by a very soft component
near peak, reasonably well fit by thermal (black-body-like) emission \citep[e.g.][]{KomossaBade,
Komossa1242,Esquej08,Miller,Lin17}. Others, such as SDSS J095209.56+214313.3, Swift 1644+57, \msevenns, 2MASS 0619-65, and
IGR J12580+0134, displayed a hard power law \citep{Komossa08,Burrows11,Saxton17,Saxton14,Walter13}
of slope $\Gamma\sim1.3-2.3$. \srcnamens, however, has a steep power law ($\Gamma\sim2.6$) that 
does not change, while the flux drops by a factor of 100\footnote{There is evidence for a deficit of high-energy photons in the later Swift observations. This had disappeared by the time of the last \xmm observation.}.
We  constrained the radio emission in this TDE to $<10\mu Jy$, the deepest upper limit for a TDE to date, which makes it unlikely that the X-rays were produced
by a jet.  The total energy of any off-axis jet can be constrained to $E<5\times10^{50}$
 ergs for circumnuclear densities $>0.1$ cm$^{-3}$ \citep{Generozov}.

The event may be self-consistently explained if the debris formed a 
viscously evolving, truncated accretion disc that emitted thermally in the UV and 
EUV bands, and a population of warm electrons that upscattered the thermal photons 
into a pseudo-power law of X-ray photons.

\section{Summary}

A flare was detected from the galaxy \srcnamem on August 22, 2016,
reaching a bolometric luminosity, $L_{bol}\sim10^{43}$ \lumUnitsns.
The source flux subsequently decayed by a factor of 100
in the 0.2--2 keV X-ray band over 600 days, but barely changed in the UV filters
for the first 400 days and then dropped by about one magnitude in the UVM2 and UVW2 bands. The galaxy shows no signs of 
previous AGN activity, and we attribute the flare to the accretion
of debris from a star that was tidally destroyed 
during a close approach to the nuclear black hole. The apparent independence 
of the X-ray and UV light curves may point to the creation of a slowly evolving,
long-lived  accretion disc structure in this event. 
The X-ray spectrum can be modelled with a power law of $\Gamma\sim2.6$
throughout the evolution of the event (although a deficit of high-energy photons was noticed during the middle of the monitoring) and may be solely due to Compton upscattering of thermal photons from the disc.
This event is radio-quiet to a level $L_{R}<4\times10^{36}$\lumunitsns, which 
constrains the total energy emitted by an off-axis jet to $<5\times10^{50}$ ergs, the
deepest limit yet achieved for an X-ray selected TDE.

\acknowledgements
We thank the anonymous referee for the useful comments which improved the manuscript,
and the XMM OTAC for approving this programme.
The XMM-Newton project is an ESA science mission with instruments and contributions directly funded by ESA member states and the USA (NASA).
The \xmm project is supported by the Bundesministerium f\"{u}r Wirtschaft 
und Technologie/Deutches Zentrum f\"{u}r Luft- und Raumfahrt i
(BMWI/DLR, FKZ 50 OX 0001), the Max-Planck Society, and the Heidenhain-Stiftung.
We thank the \swift team for approving and performing the monitoring 
observations. This work made use of data supplied by the UK \swift Science Data Centre at 
the University of Leicester. This paper uses data products produced by the OIR Telescope Data Center, supported by the Smithsonian Astrophysical Observatory.
The Liverpool Telescope is operated on the island of La Palma by Liverpool John 
Moores University in the Spanish Observatorio del Roque de los Muchachos of 
the Instituto de Astrofisica de Canarias with financial support from the UK 
Science and Technology Facilities Council.
Some of the data presented in this paper were obtained from the Mikulski Archive for Space Telescopes (MAST). STScI is operated by the Association of Universities for Research in Astronomy, Inc., under NASA contract NAS5-26555. K.D.A. acknowledges support provided by NASA through the NASA Hubble Fellowship grant HST-HF2-51403.001 awarded by the Space Telescope Science Institute, which is operated by the Association of Universities for Research in Astronomy, Inc., for NASA, under contract NAS5-26555. The National Radio Astronomy Observatory is a facility of the National Science Foundation operated under cooperative agreement by Associated Universities, Inc.
Nick Stone and Thomas Wevers are warmly thanked for helpful comments.



\end{document}